%% file: PulledStars2d_v10.tex
\documentclass{iopart}
\usepackage{iopams,amsfonts,amssymb,amsthm} 
\usepackage{geometry}                
\usepackage{etex}
\usepackage[pdftex]{graphicx}
\usepackage{rawfonts}
\usepackage{amssymb}
\usepackage{epstopdf}
\usepackage[usenames,dvipsnames,svgnames]{xcolor}
\expandafter\let\csname equation*\endcsname\relax
\expandafter\let\csname endequation*\endcsname\relax
\usepackage{amsmath}
\newcommand{\avm}{\langle v \rangle}
\newcommand{\avh}{\langle h \rangle}
\newcommand{\avz}{\langle z_0 \rangle}

\input prepictex.tex
\input pictex.tex
\input postpictex.tex

\newtheorem{lemm}{Lemma}

\def\Pr{\noindent \emph{Proof: }}
\def\qed{$\Box$}

\def\nor{\normalsize}

\def\sfrac#1#2{\hbox{\nor $\frac{#1}{#2}$}}
\def\Sfrac#1#2{\hbox{\large $\frac{#1}{#2}$}}

\usepackage{hyperref}
\hypersetup{colorlinks,citecolor=blue,filecolor=blue,linkcolor=blue,urlcolor=blue}

\begin{document}

\title{Force-induced desorption of $3$-star polymers in two dimensions}
\author{C J Bradly$^*$, E J Janse van Rensburg$^{\dagger}$, A L Owczarek$^*$ and S G  Whittington$^\ddagger$ }
\address{
{}$^\dagger$School of Mathematics \& Statistics, University of Melbourne, Victoria 3010, Australia \\
{}$^\dagger$Department of Mathematics \& Statistics, York University, Toronto M3J 1P3, Canada \\
{}$^\ddagger$Department of Chemistry, University of Toronto, Toronto M5S 3H6, Canada 
}

\begin{abstract}
We investigate the phase diagram of a self-avoiding walk model of a 3-star polymer in two dimensions, adsorbing at a surface and being desorbed by the action of a force.  
We show rigorously that there are four phases: a free phase, a ballistic phase, an adsorbed phase and a mixed phase where part of the 3-star is adsorbed and part is ballistic.  
We use both rigorous arguments and Monte Carlo methods to map out the phase diagram, and investigate the location and nature of the phase transition boundaries.
In two dimensions, only two of the arms can be fully adsorbed in the surface and this alters the phase diagram when compared to 3-stars in three dimensions.
\end{abstract}

\maketitle

\section{Introduction}
\label{sec:Introduction}
\label{sec:introduction}

There has been considerable recent interest in how self-avoiding walks 
\cite{Rensburg2015,MadrasSlade} respond to a force.   For a review see \cite{Orlandini}. 
A particularly interesting situation is when the walk is adsorbed at a surface and 
is then desorbed by a force applied at the last vertex of the walk
\cite{Guttmann2014,Rensburg2013,Krawczyk2005,Krawczyk2004,Mishra2005}.  
For related work see for instance \cite{Skvortsov2009} and \cite{Binder2012}.

It is natural to ask how the architecture of a polymer affects its adsorption properties
and how it responds to a force. This was investigated for lattice polygons 
(a model of ring polymers) \cite{Guttmann2018} and we have a fairly 
complete understanding of the behaviour in three dimensions.  In two dimensions 
the situation is  more difficult but Beaton \cite{Beaton2017} has given an essentially 
complete solution for staircase polygons in 
two dimensions.  Various models of branched polymers have also been
investigated \cite{Bradly2019,Rensburg2018,Rensburg2019}.

\begin{figure}[b!]
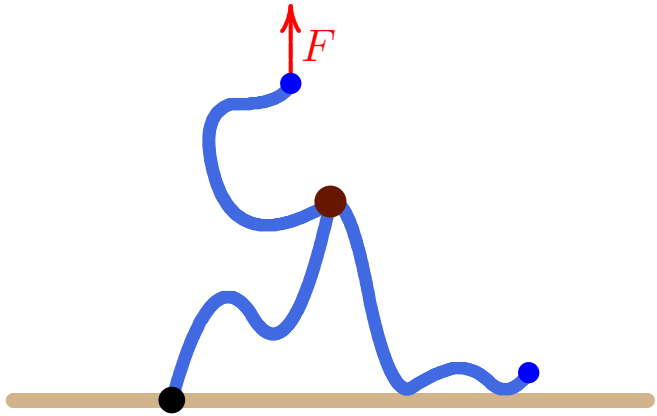

\centering
\input figure01.tex
\label{figure01}
\caption{A pulled adsorbing 3-star polymer.}
\end{figure}

In this paper we shall be concerned with pulled adsorbing star polymers (see figure
\ref{figure01}).
A \emph{star} with $f$ arms, or an \emph{$f$-arm star}, is a connected
graph with no cycles, one vertex of degree $f$ and $f$ vertices of degree $1$.   The
star is \emph{uniform} if all the arms have the same number of edges.  In this paper
we shall only be concerned with the uniform case and we shall drop the adjective unless
it is likely to cause confusion.  
We shall count embeddings in the $d$-dimensional hypercubic lattice, $\mathbb{Z}^d$,
of stars with one vertex of degree $1$ fixed at the origin.
Write $s_n^{(f)}$ for the number of such embeddings 
with a total of $n$ edges.  Note that $f$ must divide $n$.  For the $d$-dimensional 
hypercubic lattice $\mathbb{Z}^d$ with $f=3, \ldots, 2d$, we know that 
\cite{WhittingtonSoteros1991,WhittingtonSoteros1992}
\begin{equation}
\lim_{n\to\infty} \sfrac{1}{n} \log s_n^{(f)} = \log \mu_d
\label{eqn:stargrowth}
\end{equation}
where $\mu_d$ is the growth constant of the $d$-dimensional 
hypercubic lattice (and the limit $n\to\infty$ is taken through $n=fm$
(multiples of $f$ in $\mathbb{N}$)).  

To investigate how an adsorbed star responds to a force we can attach the 
star to an impenetrable surface at a vertex of degree 1 and pull at another vertex of 
degree 1.  By keeping track of how many vertices are in the surface and the 
height of the vertex where the force is applied we can map out the phase
diagram of the system.  This has been carried out for 3-stars in three 
dimensions both rigorously \cite{Rensburg2018} and using Monte Carlo
methods \cite{Bradly2019}.  In three dimensions we have a considerable amount of
information available about the phase diagram \cite{Bradly2019,Rensburg2018}.
In two dimensions the situation seems more complicated since it is not possible
for the star to lie entirely in the surface at which adsorption occurs.  Instead, in
the adsorbed phase two arms may be adsorbed and screen the third arm from
interacting with the adsorbing surface.  This changes the character of the
adsorbed phase and makes rigorous treatment of the model difficult.   We augment
our results by Monte Carlo simulations to map out the phase diagram of the model. 
We rigorously identify four phases, namely a free phase, and  ballistic, adsorbed,
and mixed phases.  We rigorously determine three of the phase boundaries
between these phases, and give numerical evidence for the location of the
fourth (adsorbed-mixed) phase boundary.

\section{A brief review}
\label{sec:review}

This section gives some rigorous results about self-avoiding walks
adsorbed at a surface and subject to a force that can desorb the walk.  We shall
need some of these result in Section \ref{sec:rigorous}.

Consider the $d$-dimensional hypercubic lattice $\mathbb{Z}^d$.  The vertices in
this lattice have coordinates $(x_1,x_2, \ldots x_d)$, $x_i \in \mathbb{Z}$.
Suppose that $c_n$ is the number of $n$-edge self-avoiding walks starting at the origin.
Then \cite{Hammersley1957}
\begin{equation}
\log d \le \lim_{n\to\infty} \sfrac{1}{n} \log c_n = \log \mu_d \le \log(2d-1)
\end{equation}
where $\mu_d$ is the growth constant of the self-avoiding walk.  Note that 
the numbers of 
self-avoiding walks and uniform stars grow at the same exponential rate \cite{WhittingtonSoteros1991}.
If the walk is constrained to lie in or on one side of the hyperplane
$x_d=0$ we call the walk a 
\emph{positive walk} and write $c_n^+$ for the number of $n$-edge
positive walks.    It is known \cite{Whittington1975} that
\begin{equation}
\lim_{n\to\infty} \sfrac{1}{n} \log c_n^+ = \log \mu_d.
\end{equation}

Suppose that $c_n^+(v,h)$ is the number
of $n$-edge positive walks with $v+1$ vertices in the hyperplane 
$x_d=0$ and with the $x_d$-coordinate of the last vertex equal to 
$h$.  We say that the walk has $v$ \emph{visits} and the last vertex
has \emph{height} equal to $h$.  Define the partition function
\begin{equation}
C_n^+(a,y) = \sum_{v,h} c_n^+(v,h) a^v y^h,
\end{equation}
where $a=\exp(\epsilon/k_\text{B}T)$ and $y=\exp(F/k_\text{B}T)$ are the Boltzmann weights associated with the monomer-surface interaction energy $-\epsilon$ and the pulling force $F$, respectively.

Suppose that the positive walk interacts with the surface but is not subject to a force
(so that $y=1$).  Then the (reduced) free energy is 
\begin{equation}
\kappa(a) = \lim_{n\to\infty} \sfrac{1}{n} \log C_n^+(a,1)
\end{equation}
and we know that there exists a critical value of $a$, $a_c > 1$,
such that $\kappa(a) = \log \mu_d$ when $a \le a_c$ and 
$\kappa(a) > \log \mu_d$ when $a > a_c$.  The free energy $\kappa(a)$ is
singular at $a=a_c > 1$ \cite{HTW,Rensburg1998,Madras} and it is
a convex function of $\log a$ \cite{HTW}.

If the walk is subject to a force but does not interact
with the (impenetrable) surface then $a=1$ and the free energy is
\begin{equation}
\lambda(y) = \lim_{n\to\infty} \sfrac{1}{n} \log C_n^+(1,y).
\end{equation}
$\lambda(y)$ is singular at $y=1$ \cite{Beaton2015,IoffeVelenik,IoffeVelenik2010} and
the walk is in a ballistic phase when $y > 1$.  Also $\lambda(y)$
is a convex function of $\log y$ \cite{Rensburg2009}.

If we return to the general situation where $a \ne 1$ and $y \ne 1$ then the 
limit defining the free energy exists \cite{Rensburg2013} and the free energy is given by
\begin{equation}
\psi(a,y) = \lim_{n\to\infty} \sfrac{1}{n} \log C_n^+(a,y).
\end{equation}
Moreover, it is known \cite{Rensburg2013} that
\begin{equation}
\psi(a,y) = \max[\kappa(a), \lambda(y)]
\label{eqn:psicondition}
\end{equation}
and, in particular, $\psi(a,y) = \log \mu_d$ when $a \le a_c$
and $y \le 1$.  For $a > a_c$ and $y > 1$ there is a phase
boundary in the $(a,y)$-plane along the curve 
given by $\kappa(a) = \lambda(y)$.  This phase transition
is first order \cite{Guttmann2014}.

Several homeomorphism types (corresponding to different 
polymer architectures) have been investigated including 
polygons (as a model of ring polymers) \cite{Guttmann2018} and various
types of branched polymers \cite{Bradly2019,Rensburg2018,Rensburg2019}.
In particular, consider 3-star polymers modelled as 3-stars on the simple cubic lattice
\cite{Rensburg2018}.
The 3-star is terminally attached to an impenetrable surface at which it can 
adsorb, and pulled normal to the surface at another unit degree vertex.  The free energy
has been shown \cite{Rensburg2018}  to be given by the expression
\begin{equation}
\sigma^{(3)}(a,y) = \max \left[ \Sfrac{1}{3}(2\lambda(y)+\log \mu_3), 
\Sfrac{1}{3}(\lambda(y) + 2 \kappa(a)), \kappa(a)  \right].
\end{equation}
Each of the terms in this expression corresponds to a phase, so that we have
ballistic, mixed and adsorbed phases, in addition to a free phase with free energy equal to 
$\log \mu_3$.  This model has also been studied using a Monte Carlo 
approach \cite{Bradly2019} and all these phases, and the corresponding phase
boundaries, are clearly seen in that study.

\section{Some rigorous results}
\label{sec:rigorous}

In \cite{Rensburg2018} we considered the case of a 3-star on the simple
cubic lattice $\mathbb{Z}^3$, terminally attached to a surface, and pulled 
at another unit degree vertex.  Many of the arguments in that paper work on 
the two-dimensional square lattice $\mathbb{Z}^2$, but with an 
important exception.  When we consider the adsorbed phase of 
the 3-star, it is not possible for all the vertices of the star to be 
adsorbed in the surface because two arms of  the star partially shade the 
surface from the third arm.  This results in a problem 
with fully characterizing the free energy of the adsorbed phase.  We can construct a lower bound on the free energy corresponding to two arms being in the surface and the third arm being out of the surface, but we cannot construct a corresponding upper bound.  As we shall see,
however, it is still possible to make some useful predictions about the form
of the phase diagram in two dimensions.

We consider $\mathbb{Z}^2$ with the obvious coordinate
system $(x_1,x_2)$ so that all vertices have integer coordinates.  We
consider uniform 3-stars with a vertex of degree 1 at the origin, all vertices having
non-negative $x_2$-coordinate, pulled (in the $x_2$-direction)
at another vertex of degree 1.  Suppose
that the (uniform) star has $n$ edges, with $n$ being a multiple
of 3 so that each arm has $n/3$ edges.  We write $s_n^{(3)}(v,h)$ for the
number of these stars with $n$ edges, $v+1$ vertices in the surface $x_2=0$ and with the $x_2$-coordinate of the unit degree vertex  where the 
force is applied equal to $h$.  We call  $v$ the number of \emph{visits}
and $h$ the \emph{height}.  The partition function is given by 
\begin{equation}
S_n^{(3)}(a,y) = \sum_{v,h} s_n^{(3)}(v,h) a^v y^h.
\label{eqn:partitionfunction}
\end{equation}

Our aim is to find useful bounds on the free energy.  We obtain lower bounds by
strategy arguments and these arguments are essentially those used in 
\cite{Rensburg2018}.  To obtain upper bounds we consider the arms of the 
star as independent and consider cases where one, two or three arms have 
vertices in $x_2=0$.

We state the lower bounds as a Lemma:
\begin{lemm}
For all values of $a$ and $y$
$$\liminf_{n\to\infty} \sfrac{1}{n} \log S_n^{(3)}(a,y) \ge \max\left[  
\Sfrac{1}{3}(2\lambda(y) + \log \mu_2), 
\Sfrac{1}{3}(\lambda(y)+2\kappa(a)) \right].$$
\label{lem:lowerbound}
\end{lemm}
The proof of Lemma \ref{lem:lowerbound} is essentially the same as the 
corresponding proof given in \cite{Rensburg2018} and we do not repeat it here.
The first term corresponds to the ballistic phase where two arms are pulled and the third
is free, while the second term corresponds to a mixed phase with two arms adsorbed and 
the third being pulled.  We show that each of these bounds is sharp in certain regions
of the phase diagram.
Note however that in three dimensions there is an additional lower bound of 
$\kappa(a)$, corresponding to all three arms being adsorbed (see lemma 6 in reference
\cite{Rensburg2018}).  In three dimensions one 
can confine all three arms to disjoint wedges each of which allows adsorption.
In two dimensions the partial screening of the surface by two arms makes this
not possible.  

The best upper bound that we have is as follows:
\begin{lemm}
For all values of $a$ and $y$
$$\limsup_{n\to\infty} \sfrac{1}{n} \log S_n^{(3)}(a,y) \le \max
\left[\Sfrac{1}{3}(2\lambda(y) + \log \mu_2), 
\Sfrac{1}{3}(\lambda(y)+2\kappa(a)),\kappa(a) \right].$$
\label{lem:upperbound}
\end{lemm}
\Pr
If only one arm has vertices in $x_2=0$ the free energy is bounded above by
$$\Sfrac{1}{3}(\max[\kappa(a), \lambda(y)]+ \lambda(y) + \log \mu_2).$$  
To obtain this treat the three 
arms as independent.  Then the contributions from the three arms are the three
terms in the above expression.  If two arms have vertices in the surface their 
contribution to the free energy is $\max[2\lambda(\sqrt{y}), 2\kappa(a)]$ 
\cite{Rensburg2017} while
the third arm contributes $\lambda(y)$.  Hence the free energy is bounded above
by
$$\Sfrac{1}{3}(\max[2\lambda(\sqrt{y}), 2 \kappa(a)]+\lambda(y))
\le  \Sfrac{1}{3}(\max[\lambda(y) + \log \mu_2, 2 \kappa(a)]+\lambda(y)),$$
where this inequality follows from the log convexity of $\lambda(y)$ \cite{Rensburg2009}.
If all three arms have vertices in $x_2=0$ the free energy is bounded above by
$$\Sfrac{1}{3}(\max[\lambda(y),\kappa(a)] +2 \kappa(a)).$$  
Recall that $\kappa(a) \ge \log \mu_2$.
Putting these upper bounds together  completes the proof.
\qed

If $y \le 1$ $\lambda(y) = \log \mu_2$ and if $a \le a_c$ $\kappa(a) = \log \mu_2$.
Therefore if $y \le 1$ and $a \le a_c$ the lower and upper bounds all give $\log \mu_2$
so $\sigma^{(3)}(a,y) = \lim_{n\to\infty} n^{-1} \log S_n^{(3)}(a,y) = \log \mu_2$ and the system is in the \emph{free phase}.

If $y > 1$ and $\lambda(y) >   2\kappa(a) - \log \mu_2$ the free energy is given by
\begin{equation}
\sigma^{(3)}(a,y) = \Sfrac{1}{3}(2 \lambda(y) + \log \mu_2)
\label{eqn:FEballistic}
\end{equation}
and the system is in the \emph{ballistic phase}.  Since $\lambda(y)$ is singular at $y=1$ 
\cite{Beaton2015}
there is a phase boundary between the free phase and the ballistic phase 
at $y=1$ for $a < a_c$.

If $2\kappa(a) - \log \mu_2 \ge \lambda(y) \ge \kappa(a)$ the free energy is given by
\begin{equation}
\sigma^{(3)}(a,y) = \Sfrac{1}{3}(\lambda(y) + 2\kappa(a)).
\label{eqn:FEmixed}
\end{equation}
The free energy depends on both $a$ and $y$ and we say that the system is in 
a \emph{mixed phase}.  There is a phase boundary between the ballistic and mixed
phases at the solution of $\lambda(y) = 2\kappa(a) - \log \mu_2$.  Since 
$\lambda(y)$ is convex in $\log y$ and $\kappa(a)$ is convex in $\log a$, they
are differentiable ae. and strictly increasing if $y>1$ and $a>a_c$.  This shows
that $\lambda(y)$ has an inverse function $\lambda^{-1}(x)$ which is continuous
and increasing for $x>\log \mu_2$ and is also differentiable ae.  The phase boundary 
between the ballistic and mixed phases is given by 
$y^{I}(a) = \lambda^{-1} \left(  2\kappa(a) - \log \mu_2\right)$ which is
continuous for all $a>a_c$ and is an increasing function (since $\lambda^{-1}(a)$ is 
increasing and continous).  The phase boundary $y^{I}(a)$ is also differentiable ae.
The methods of \cite{Guttmann2014} can be used to establish that the phase 
boundary $y^{I}(a)$ is first order, except perhaps at $(a_c,1)$.

\begin{figure}[t!]
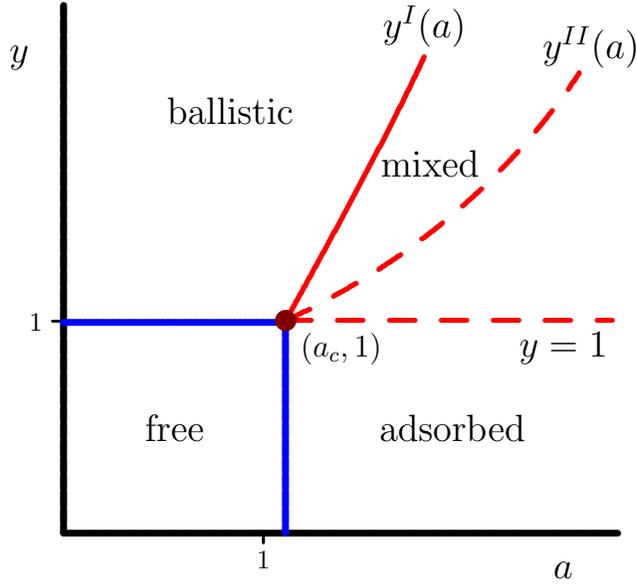

\beginpicture
\setcoordinatesystem units <2.1pt,2pt>
\setplotarea x from -40 to 100, y from -10 to 100

\color{black}
\setplotsymbol ({$\cdot$})
\plot -2 40 0 40 /   \plot 36 -2 36 0 /

\setsolid

\setplotsymbol ({\tiny$\bullet$})
\plot 0 100 0 0 100 0 /

\color{black}
\put {\Large$y$} at -8 90
\put {\Large$a$} at 90 -7
\put {$1$} at -5 40
\put {$1$} at 36 -5
\put {\Large$y^{I}(a)$} at 65 96
\put {\Large$y^{II}(a)$} at 95 92
\put {\large$(a_c,1)$} at 50 35
\put{\Large$y=1$} at 90 35

\put {\Large\hbox{free}} at 20 20 
\put {\Large\hbox{ballistic}} at 30 80 
\put {\Large\hbox{adsorbed}} at 70 20 
\put {\Large\hbox{mixed}} at 66 70

\color{blue}
\plot 40 0 40 40 0 40 /

\setplotsymbol ({\LARGE$\cdot$})
\color{red}
\setquadratic
\plot 40 40 55 69 65 90 /
\setdashes <10pt>
\plot 40 40 70 40 100 40 /
\plot 40 40 70 60 93 87 /
\setlinear
\color{Maroon}
\put {\huge$\bullet$} at 40 40

\color{black}
\normalcolor
\endpicture
\caption{The phase diagram of pulled adsorbing 3-stars in the
square lattice.  For $y\leq 1$ and $a\leq a_c$ the free energy is 
equal to $\log \mu_2$.  This is a free phase with phase boundaries 
at $y=1$ separating it from the ballistic phase, and at $a=a_c$
separating it from the adsorbed phase.  If $y> \max\{1,y^{I}(a)\}$,
then the 3-stars are in a ballistic phase with free energy given by
$(2\lambda(y) + \log \mu_2)/3$.   If $2\kappa(a)-\log \mu_2
\geq \lambda(y)\geq\kappa(a)$ then the system is in a mixed phase.
If $y\leq 1$ and $a>a_c$ then the 3-stars are adsorbed.  If
$a>a_c$ and $\kappa(a)>\lambda(y)$ then the free energy is bounded
by equations \eqref{eqn13} and \eqref{eqn14}.  Therefore, there is
a phase boundary between the solution $y^{II}(a)$ of $\lambda(y)
= \kappa(a)$ and the line $a>a_c$ and $y=1$ indicated by broken
lines.}
\label{figure02}
\end{figure}

When $\kappa(a) > \lambda(y)$ we have less information but we know that 
\begin{equation}
\liminf_{n\to\infty} \sfrac{1}{n} \log S_n^{(3)}(a,y) \ge 
\Sfrac{1}{3}(2\kappa(a) + \lambda(y))
\label{eqn13}
\end{equation}
and 
\begin{equation}
\limsup_{n\to\infty} \sfrac{1}{n} \log S_n^{(3)}(a,y) \le \kappa(a).
\label{eqn14}
\end{equation}
When $y \le 1$ we know that $\lambda(y) = \log \mu_2$.  Since 
$\kappa(a) = \log \mu_2$ when $a \le a_c$ and $\kappa(a) > \log \mu_2$
when $a > a_c$ there is a phase boundary at $a=a_c$ for $y < 1$.

When $a > a_c$, there is a phase boundary between the solution
to the equation $\lambda(y) = \kappa(a)$ (given by $y^{II}(a) = \lambda^{-1}(\kappa(a))$)
and the line $y=1$.

These results establish the locations of three phase boundaries and give bounds
on the location of a fourth boundary.  These four phase boundaries meet at 
$(a_c,1)$ which is a multicritical point in the phase diagram.

Our inability to locate the phase boundary between the adsorbed and 
mixed phases stems from the weak upper bound, $\kappa(a)$.  We expect 
that the free energy in the adsorbed phase will be $(2\kappa(a) + \log \mu_2)/3$,
corresponding to two arms being adsorbed and the third arm (partially shielded
from the surface) contributing the free energy of a free arm.
We have been unable to prove this because we cannot construct an argument giving a sharp
upper bound, in contrast with the three dimensional case (see lemma 11 in
reference \cite{Rensburg2018}).  If the free energy in the adsorbed phase was 
$(2\kappa(a) + \log \mu_2)/3$ then
the phase boundary
between the adsorbed and mixed phases would be at $y=1$ for $a > a_c$.  We
shall provide Monte Carlo evidence for this.

\section{Monte Carlo Results}
\label{sec:montecarlo}

We have simulated pulled and adsorbing uniform 3-stars on the square lattice with arm lengths up to $128$ using the flatPERM algorithm \cite{Prellberg2004}. 
The 3-stars are modelled as three self-avoiding walks grown from the origin with the surface defined as the smallest $x_2$ value of any vertex in the 3-star.
Further details of the application of flatPERM to $f$-stars are described in Ref.~\cite{Bradly2019} where it was used for 3-stars on the simple cubic lattice.
While the change from three to two dimensions introduces a difficulty in rigorously proving the free energy and phase boundaries, for Monte Carlo simulations it is trivial to change the lattice upon which the $f$-stars are embedded.
Numerical results thus complement the rigorous treatment discussed in Section \ref{sec:rigorous}.
In this section we present the numerical results in terms of the arm length $l=n/3$.

The output of the simulation are the weights $W_{nvh}$ that approximate the counts $s_n^{(3)}(v,h)$ used to construct the partition function Eq.~\eqref{eqn:partitionfunction}.
Then we calculate the order parameters $\avm/l$, $\avh/l$ and $\avz/l$ as weighted sums
\begin{equation}
	\langle Q \rangle_n(a,y) = \frac{\sum_{v,h} Q(n,v,h) W_{nvh} a^v y^h }{\sum_{v,h} W_{nvh} a^v y^h},
    \label{eq:DoSQuantity}
\end{equation}
where $Q$ is a generic thermodynamic quantity.
Another quantity of interest is the probability distribution of the number of contacts at a given temperature and interaction strength (i.e.~given $a$ and $y$): 
\begin{equation}
	P(v) = \frac{\sum_{h} W_{nvh} a^v y^h }{\sum_{v,h} W_{nvh} a^v y^h}.
    \label{eq:DoSProbability}
\end{equation}
and similar for the probability distribution of the height of the pulled vertex, $P(h)$.
Finally, we calculate the Hessian matrix of the free energy
\begin{equation} 
	H_n =
	\begin{pmatrix}
	 \Sfrac{\partial^2 S^{(3)}_n}{\partial a^2} 	& \Sfrac{\partial^2 S^{(3)}_n}{\partial a \partial y}	\\
	 \Sfrac{\partial^2 S^{(3)}_n}{\partial y \partial a} & \Sfrac{\partial^2 S^{(3)}_n}{\partial y^2}
	\end{pmatrix}
	.
	\label{eq:Hessian}
\end{equation}
For this work we ran ten independent simulations with $3\times10^4$ iterations each and averaged the results, obtaining a total of $1.2\times10^{11}$ samples at maximum arm length $l=128$.

\begin{figure}[t!]
\centering
	\begin{tabular}{ccc}
	\includegraphics[width=0.3\linewidth]{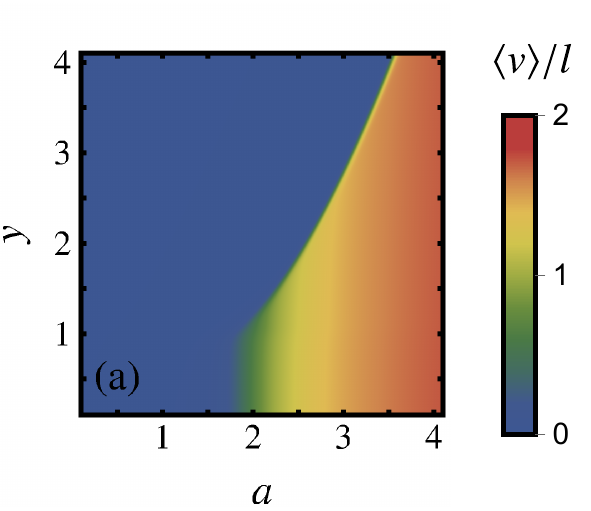}
	\label{fig:PhaseAdsorbed}
	&
	\includegraphics[width=0.3\linewidth]{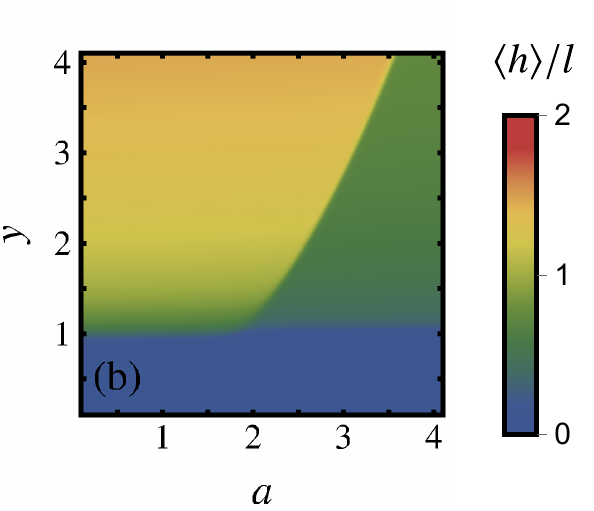}
	\label{fig:PhasePulled}
	&
	\includegraphics[width=0.3\linewidth]{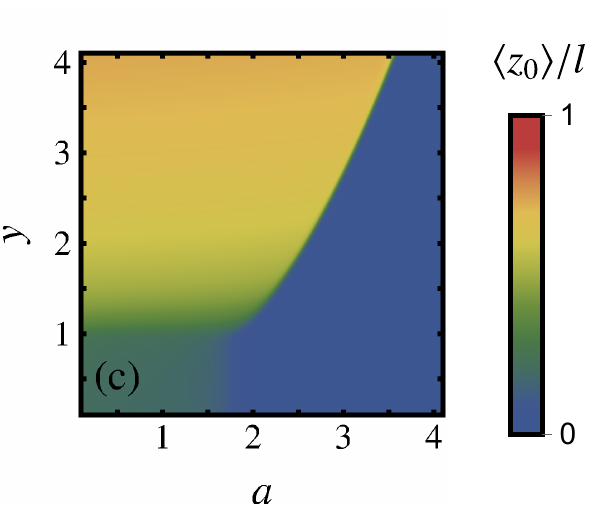}
	\label{fig:PhaseCentral}
	\end{tabular}
	\caption{The internal energies (a) $\avm/l$ and (b) $\avh/l$, and (c) height of the central vertex $\avz/l$, scaled by arm length $l=128$. 
	Four phases are apparent: the free phase for $a\leq a_\text{c}$ and $y\leq 1$; an adsorbed phase at high $a$ and small $y$; a ballistic phase at high $y$ and low $a$; and a mixed phase between the adsorbed and ballistic phases.
	}
	\label{fig:PhaseDiagram}
\end{figure}

\subsection{Phase diagram}
\label{sec:PhaseDiagram}

\begin{figure}[t!]
\includegraphics[width=\columnwidth]{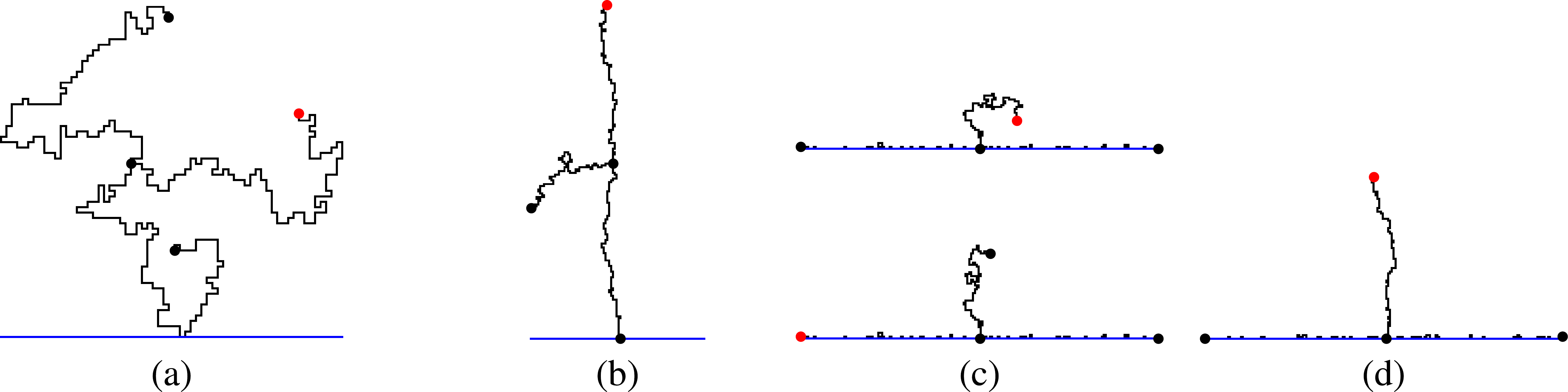}
\caption{Samples of pulled 3-stars with branch length $l=128$ taken from the Monte Carlo simulations and corresponding to likely configurations in (a) the free phase, (b) the ballistic phase, (c) the adsorbed phase and (d) the mixed phase. The force is applied at the red vertex and the blue line indicates the surface.}%
\label{fig:BestConfigurations}%
\end{figure}

To map out the phase diagram we first look at the order parameters. 
Figure \ref{fig:PhaseDiagram} shows (a) the average number of adsorbed vertices $\avm/l$, (b) the average height of the pulled vertex $\avh/l$, and (c) the average height of the central vertex $\avz/l$, each scaled by arm length $l=128$. 
Collectively, these quantities show the four phases: free, ballistic, adsorbed and mixed. In Fig.~\ref{fig:BestConfigurations} are example configurations for each phase taken from the simulations.

The free phase is bounded by the adsorption transition point at $a = a_\text{c}$ and the ballistic transition at $y = 1$, and is the region where the surface interaction is repulsive 
or insufficiently large to cause adsorption ($a \le a_\text{c}$)
and the force is absent or is a local push towards the surface ($y \le 1$). 
This matches the known result for SAWs \cite{Beaton2015}.
Within this phase both the expected number of surface contacts and the scaled average height of the pulled (or pushed in this case) vertex is zero. The free energy is therefore independent of $a$ and $y$. The configuration, shown in Fig.~\ref{fig:BestConfigurations}(a), is that of three disordered coils joined at a common end vertex.

As $a$ increases while keeping $y\leq 1$ the system undergoes a transition to the adsorbed phase at a critical value $a_\text{c}>1$.
Beyond the critical point, the average number of surface contacts $\avm/l$ quickly approaches its maximum value $2$ while $\avh/l$ and $\avz/l$ are suppressed to zero.
Further, within the adsorbed phase the free energy is independent of $y$.
This indicates that two arms are adsorbed while the third is screened from the surface forming a free coil in the bulk.
However there are two configurations with the same properties, shown in Fig.~\ref{fig:BestConfigurations}(c), namely that the values of the order parameters are the same or similar.
The distinction depends on whether the screened arm is the one with the force applied to it.
This will be important when investigating the transition to the mixed phase.

Starting again in the free phase, as $y$ increases the system enters the ballistic phase at $y=1$, where the thermodynamics depends only on the pulling force.
This phase is characterised by $\avm/l$ tending to zero while $\avh/l$ and $\avz/l$ are of order $2$ and $1$, respectively.
The expected configuration is that the pulled and tethered arms are stretched out away from the surface while the third arm assumes a disordered coil configuration relative to the central vertex, see Fig.~\ref{fig:BestConfigurations}(b).
We note that even in this phase $\avh/l$ only slowly approaches its maximum of $2$ as $y$ is increased, whereas $\avz/l$ finds its maximum of $1$ more quickly.
Analogously to the adsorbed phase, now the free energy is independent of $a$.

Between the adsorbed and ballistic phases is a mixed phase where $\avm/l$ is of order $2$, $\avh/l$ is of order $1$ and $\avz/l$ vanishes.
The free energy in this phase thus depends on both $a$ and $y$.
Similar to the adsorbed phase, two arms are adsorbed but in this phase the pulled arm extends away from the surface, see Fig.~\ref{fig:BestConfigurations}(d).
In Fig.~\ref{fig:PhaseDiagram}(b) we also see the first evidence that the boundary between the adsorbed and mixed phases is at $y=1$ for all $a > a_c$.

\subsection{Phase transitions}
\label{sec:PhaseTransitions}

\begin{figure}[t!]
\centering
	\includegraphics[width=0.55\linewidth]{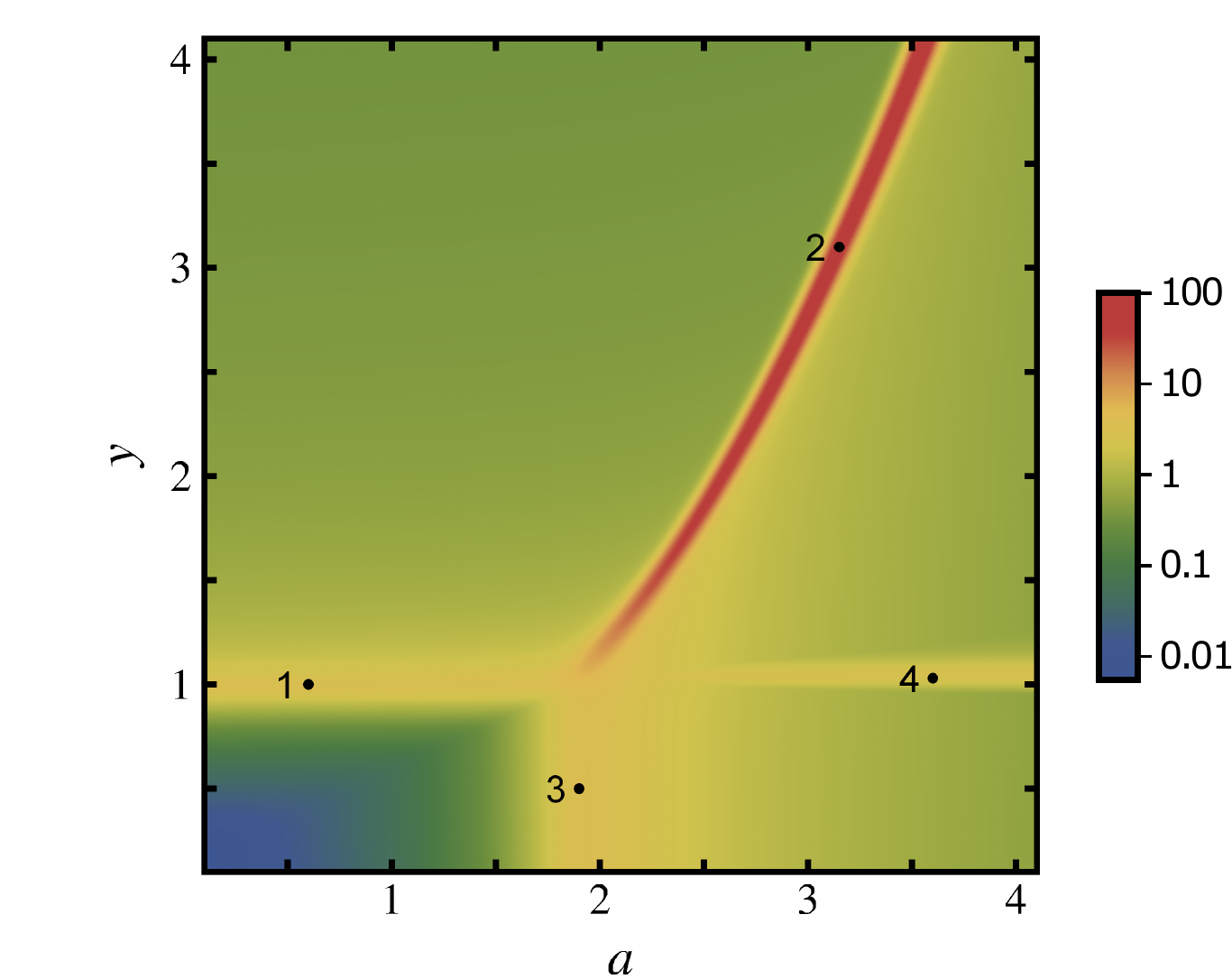}
	\vspace{0.3cm}
	\includegraphics[width=\linewidth]{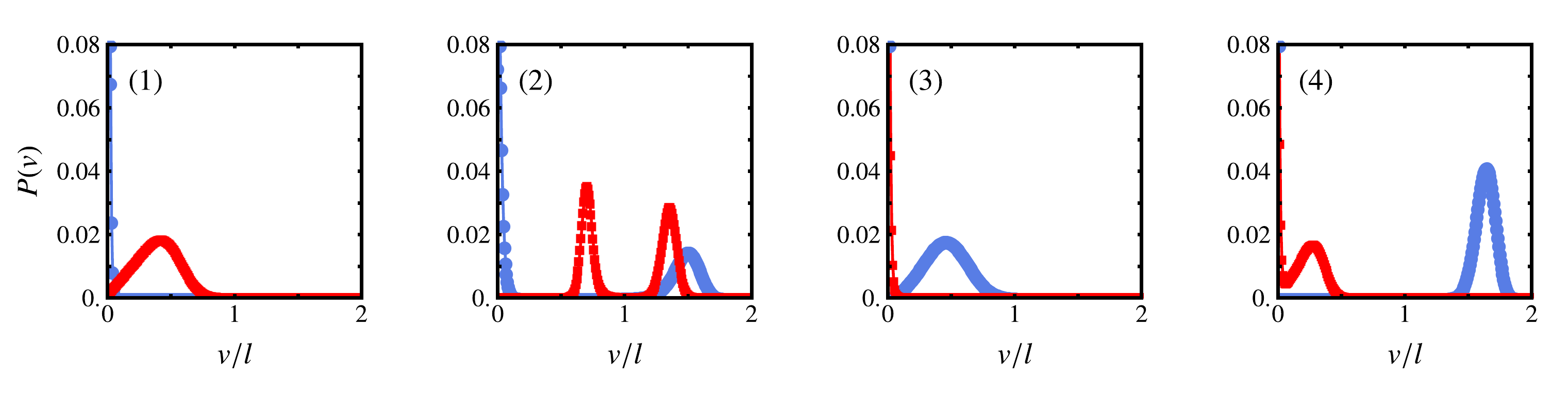}
	\caption{(Top) Density plot of the logarithm of the largest eigenvalue of the Hessian matrix of the free energy for $l=128$. 
	The phase boundaries are clearly visible as lines of high variance. 
	(Bottom) The distribution of number of adsorbed vertices (blue) and the distribution of height of pulled vertex (red) at points of interest as marked on the density plot.  
	}

	\label{fig:HessianAndDistributions}
\end{figure}

In Fig.~\ref{fig:HessianAndDistributions} we show a density plot of the logarithm of the largest eigenvalue of $H_n$ using data for $l=128$. 
The ballistic-mixed phase boundary is distinctly visible indicating a sharp and strong transition.
The ballistic and adsorbed-mixed transitions at $y=1$ are weaker yet still narrow and the adsorption transition is weaker and broad.

The type of transition is determined by looking at the underlying distributions $P(v)$ and $P(h)$ of the number of surface contacts and the height of the pulled vertex, respectively. 
The distributions at several points of interest in the $a$-$y$ plane are shown in Fig.~\ref{fig:HessianAndDistributions} and are indicative of all points along the phase boundaries.
We see that for the ballistic-mixed transition (point 2) the distributions $P(v)$ and $P(h)$ are bimodal, characteristic of a first-order transition. 
In contrast, the distributions near the free-ballistic (point 1) and free-adsorbed (point 3) boundaries are not bimodal and these transitions are continuous as expected from the case of SAWs.

The nature of the adsorbed-mixed transition (point 4) is less clear since $P(h)$ is bimodal but $P(v)$ is not. 
This matches the result from the order parameters which indicates that across the transition there are two adsorbed arms, but the non-adsorbed arm changes from the pulled arm to the free arm, as soon as the force is not towards the surface, i.e.~$y\geq1$; recall the two similar configurations in Fig.~\ref{fig:BestConfigurations}(c). 
At $y=1$, where there is no force, the arm whose endpoint height is being measured by $h$ is either adsorbed or it is screened from the surface and thus free in the bulk.
In the latter case, we expect that the height of its endpoint should scale like the end-to-end size of a free SAW in two dimensions, that is, $\avh\sim l^{3/4}$.
Although we do not have sufficient data to measure this effect, the position of the broad peak in $P(h)$ is thus expected to scale as $l^{-1/4}$.
The narrow peak in $P(h)$ at $h/l=0$ corresponds to configurations where the pulled arm is fully adsorbed so its endpoint is most likely on the surface.
Both configurations are sampled by the simulation and so $P(h)$ appears bimodal even though the screening effect means it is not a first-order transition.
Intuitively, the adsorbed-mixed transition is clear for the top configuration in Fig.~\ref{fig:BestConfigurations}(c); the force pulls the branch taut and the chain smoothly transforms to the mixed phase configuration in Fig.~\ref{fig:BestConfigurations}(d).
For the bottom configuration, transforming to the mixed phase configuration seems like a different process since in two dimensions the branches cannot move past each other and so the entire star would seem to `flip over' while preserving two adsorbed branches before the force pulls one branch taut.
This is not a concern in the thermodynamic ensemble where both configurations contribute, but the presence of both explains the properties of the adsorbed-mixed transition in comparison to the three-dimensional case where there is no distinction between probable configurations in the adsorbed phase.

\begin{figure}[t!]	
\centering
\begin{tabular}[t]{cccc}
	\includegraphics[width=0.33\linewidth]{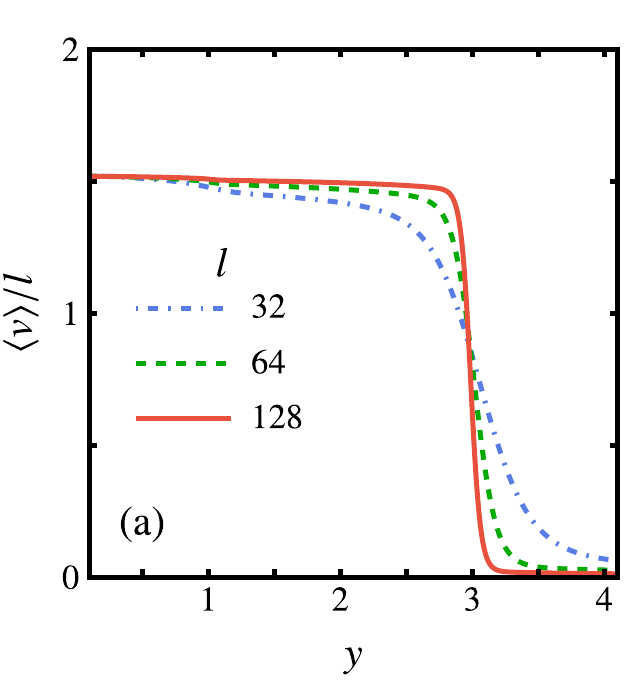}&&&
	\includegraphics[width=0.33\linewidth]{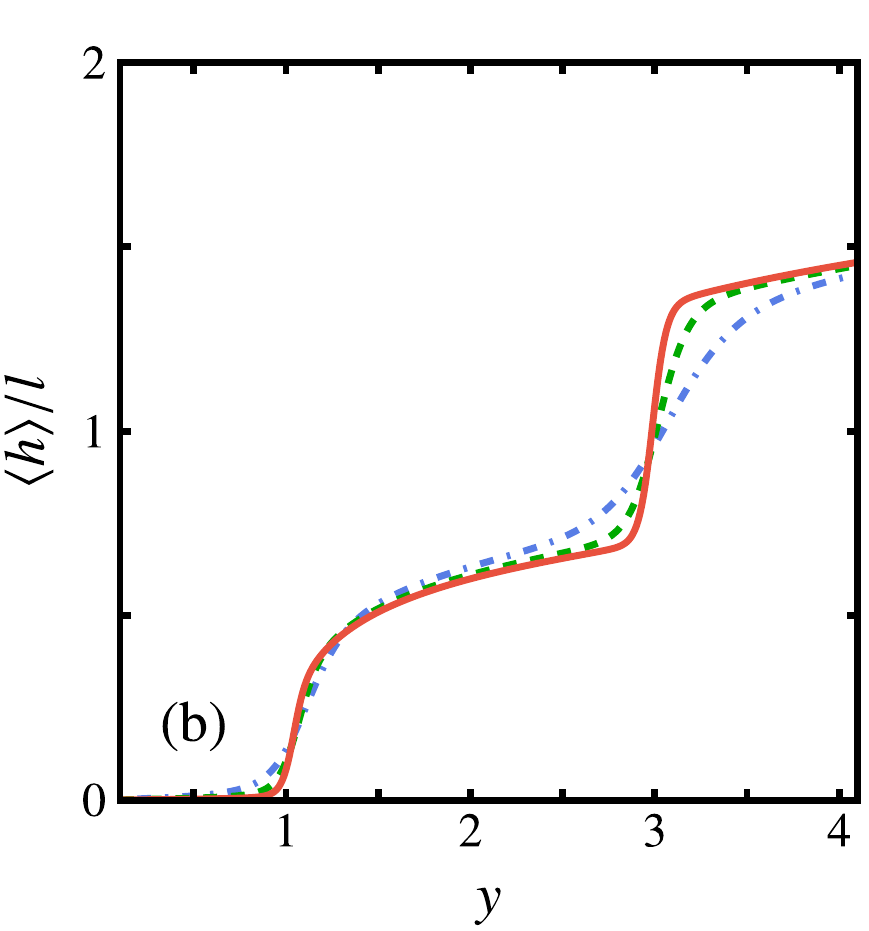}
\end{tabular}
\begin{tabular}[t]{cccc}
	\includegraphics[width=0.33\linewidth]{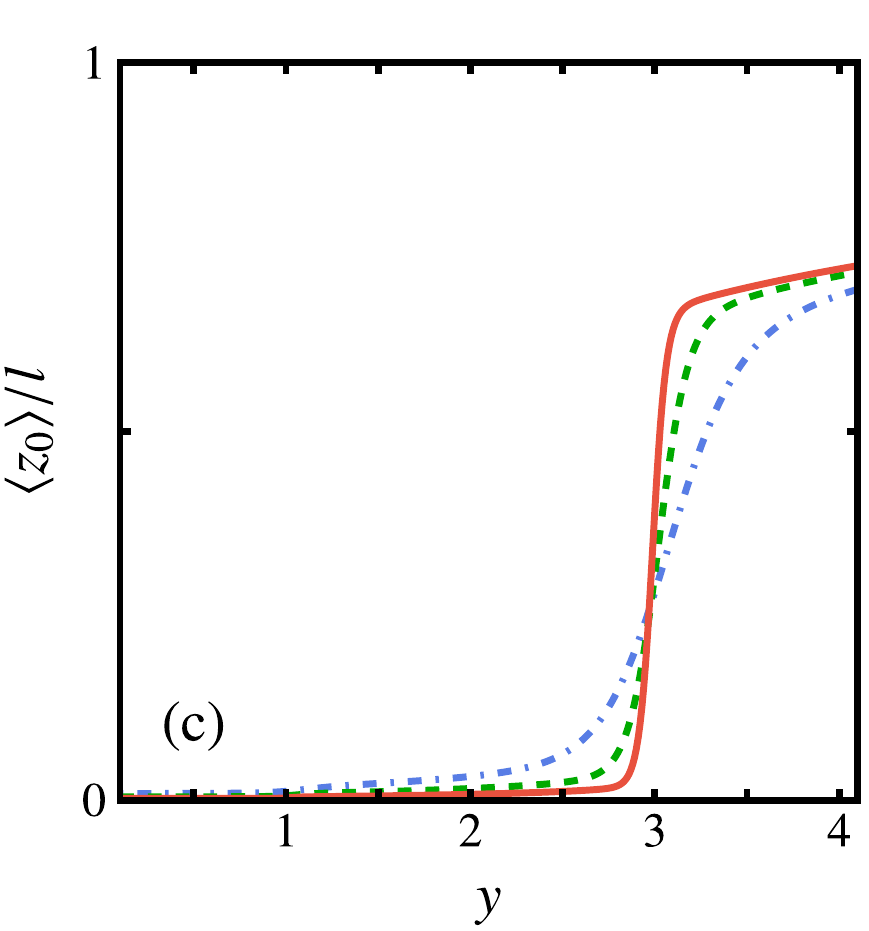}&&&
	\includegraphics[width=0.33\linewidth]{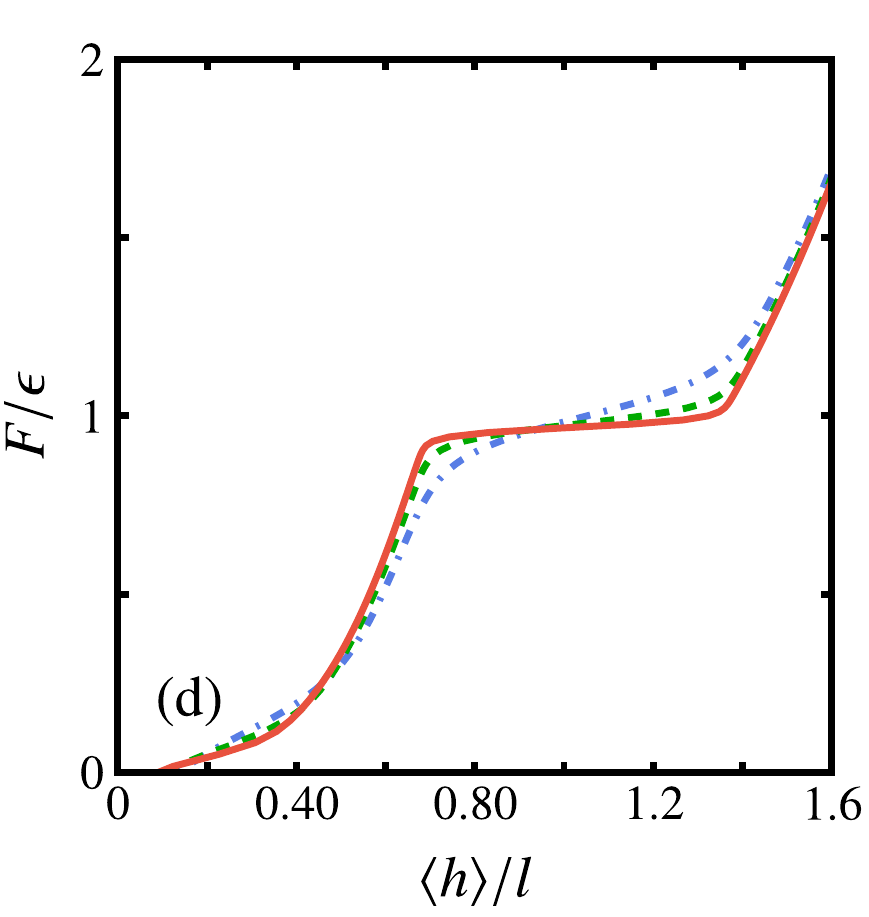}
\end{tabular}

	\caption{Internal energies (a) $\avm/l$ and (b) $\avh/l$ and (c) height of central vertex $\avz/l$ as a function of $y$ for fixed $a=3.1$ and $l=32,64,128$. 
	The ballistic-mixed transition is visible at $y\approx 3.0$ in all three quantities while and the weaker adsorbed-mixed transition at $y=1$ only affects the height of the pulled vertex $\avh$.
	(d) Force-extension graph at fixed temperature corresponding to fixed $a=3.1$.
	}

	\label{fig:EnergySlices}
\end{figure}

To confirm the nature of the ballistic-mixed and adsorbed-mixed transitions we plot in Fig.~\ref{fig:EnergySlices} the internal energies (a) $\avm/l$ and (b) $\avh/l$ as well as (c) the height of the central vertex $\avz/l$ as a function of $y$ for fixed $a=3.1$ at several values of $l$.
This is a vertical slice through the phase diagram near to points 2 and 4 in Fig.~\ref{fig:HessianAndDistributions}.
As $l$ increases, the ballistic-mixed transition ($y\approx 3.0$) appears in all three order parameters as a sharply defined latent heat, further indicating a first-order transition.
The adsorbed-mixed transition at $y = 1.0$ only appears in $\avh/l$.
While these data are only for finite-size 3-stars, as $l$ increases the singularity in $\avh/l$ looks more like a continuous transition than the discontinuous jump of a first-order transition.
The unimodal distribution $P(v)$ together with the continuous behaviour of $\avh$ indicate that the adsorbed-mixed transition is continuous in two-dimensions.
This marks a difference to the three-dimensional case where both the ballistic-mixed and adsorbed-mixed transitions are first order, showing bimodal distributions for both $P(v)$ and $P(h)$, as well as emerging latent heats for increasing $n$ \cite{Bradly2019}.

The other quantity of interest is the pulling force applied versus the extension of the polymer, as measured by atomic-force microscopy experiments, which are performed at a fixed temperature $T$ below the adsorption transition temperature \cite{Alvarez2009}.
In our parameterization this corresponds to a plot of the force $F$ (in units of $\epsilon$) versus the scaled average height of the pulled vertex $\avh/l$, where $F/\epsilon=\log y/\log a$.
In Fig.~\ref{fig:EnergySlices}(d) we show a force-extension plot at fixed temperature corresponding to $a=3.1$.
As $l$ increases we see the formation of a plateau in force $F$ as the extension is increased through the ballistic-mixed transition. It is less clear that there is a plateau forming at smaller $\langle h \rangle/l$ corresponding to the adsorbed-mixed transition.  Such a plateau would be close to $F=0$. This further suggests that the adsorbed-mixed transition is not first-order, especially in contrast to the three-dimensional case where this transition is obvious in the force-extension plane \cite{Bradly2019}.

\subsection{Phase boundaries}
\label{sec:PhaseBoundaries}

\begin{figure}[t!]
	\centering
	\begin{tabular}{cc}
	\includegraphics[width=0.45\columnwidth]{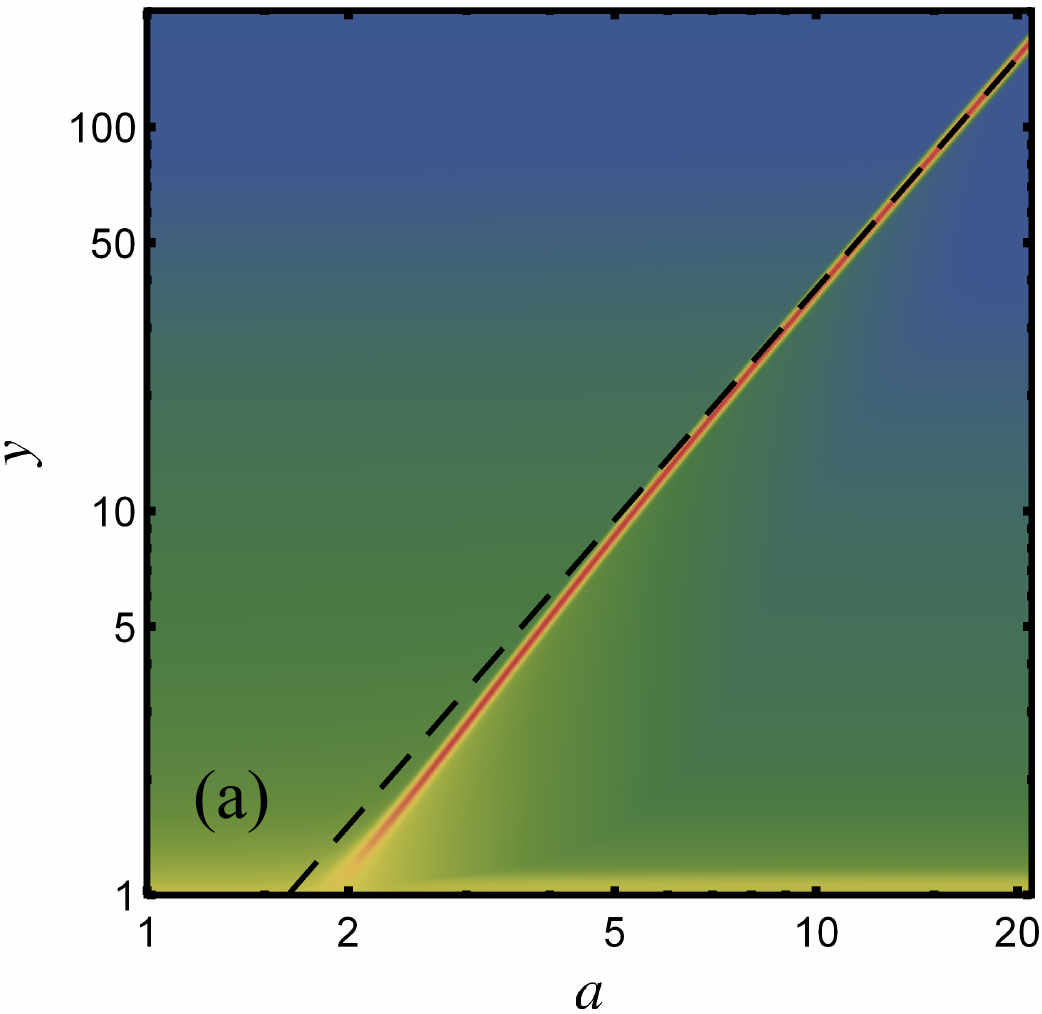}
	&
	\includegraphics[width=0.44\columnwidth]{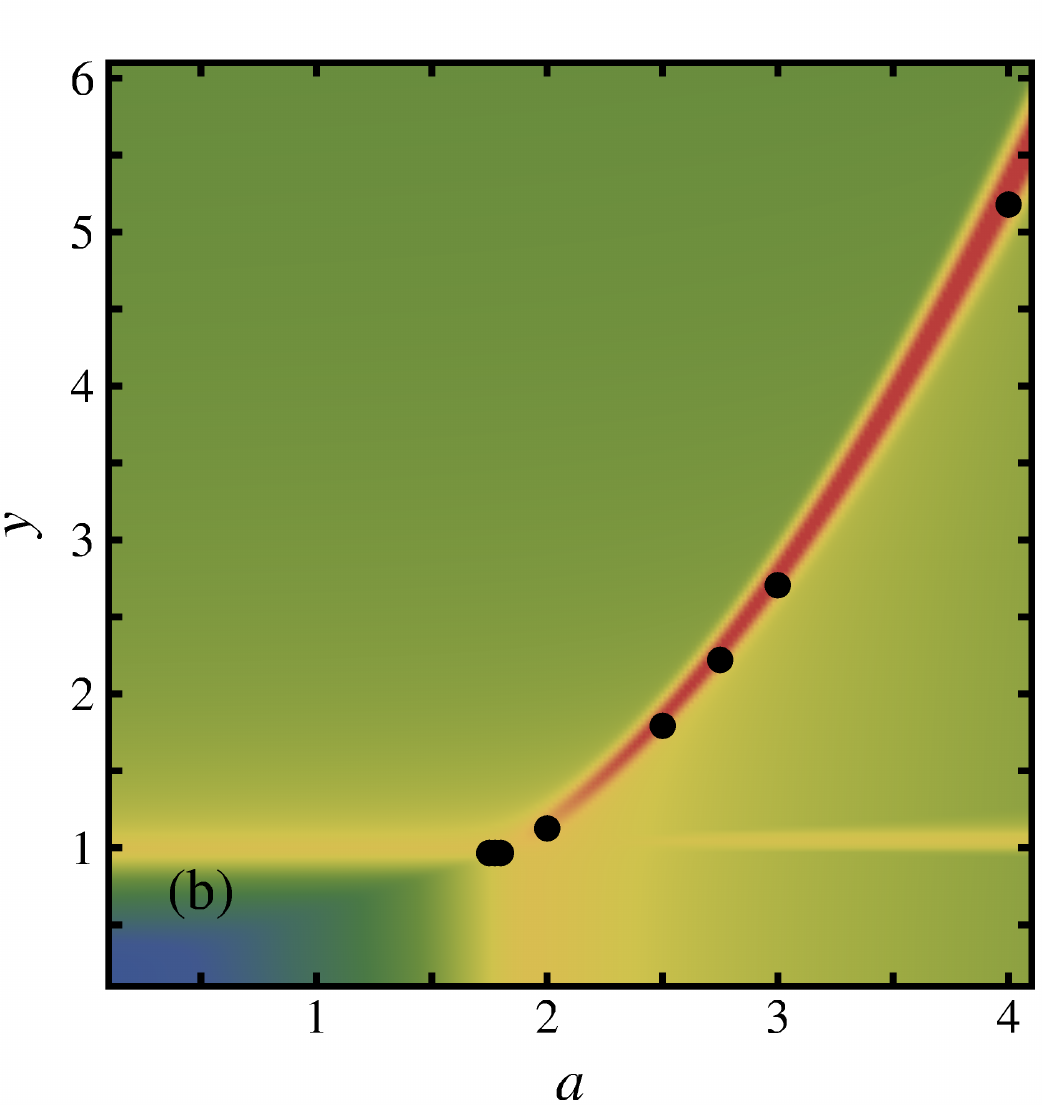}
	\end{tabular}
	\caption{The logarithm of the largest eigenvalue of the covariance matrix: (a) for large $a$ and $y$, shown on a logarithmic scale, overlaid with the asymptotic boundary of Eq.~\eqref{eq:Asymptotes} (dashed line);
	and (b) overlaid with solutions to Eq.~\eqref{eq:BallisticMixedBoundary} from exact enumeration \cite{Guttmann2014}.}%
	\label{fig:PhaseAsymptotes}%
\end{figure}

The preceding results confirm our expectation that the adsorbed-mixed boundary is at $y=1$ for all $a>a_\text{c}$.
In Section \ref{sec:rigorous} we showed that the ballistic-mixed boundary is at the solution of
\begin{equation}
	\lambda(y) = 2\kappa(a) - \log \mu_2.
	\label{eq:BallisticMixedBoundary}
\end{equation} 
We know that $\kappa \sim \log a$ for large $a$ and  $\lambda(y) \sim \log y$ for large $y$.
Thus the ballistic-mixed boundary for large $a$ and $y$ is
\begin{equation}
	y \sim \frac{a^2}{\mu_2}.
	\label{eq:Asymptotes}
\end{equation}
In Fig.~\ref{fig:PhaseAsymptotes}(a) we show the logarithm of the largest eigenvalue of $H_n$ for $l=128$ on a log-log plot for a larger range of $a$ and $y$.
Equation \eqref{eq:Asymptotes} is superimposed as a dashed line using the known value for $\mu_2$ \cite{Jensen1998}.
It is immediately clear that the ballistic-mixed boundary has the expected asymptotic form for large $a$ and $y$.

For smaller values of $a$ and $y$ Guttmann {\em et al.} have calculated the SAW free energies $\kappa(a)$ and $\lambda(y)$ to high accuracy using exact enumeration and series analysis methods \cite{Guttmann2014}.
In Fig.~\ref{fig:PhaseAsymptotes}(b) we show the logarithm of the largest eigenvalue of $H_n$ for $l=128$ overlaid with solutions to Eq.~\eqref{eq:BallisticMixedBoundary} calculated using data from Tables 1 and 2 in Ref.~\cite{Guttmann2014}.
The agreement with our Monte Carlo results for the ballistic-mixed boundary is good for $a>a_\text{c}$.


\section{Discussion}
\label{sec:discussion}

We have investigated the phase diagram of a 3-star polymer in two dimensions when the 3-star interacts with the surface where adsorption occurs, and is pulled at a vertex of degree 1. 
The problem has been studied in three dimensions both by rigorous arguments \cite{Rensburg2018} and by Monte Carlo methods \cite{Bradly2019}. 
For the two-dimensional case we have established rigorously the locations of the phase boundaries between 1) the free phase and the ballistic phase, 2) the free phase and the adsorbed phase, and 3) the ballistic phase and the mixed phase.  
For the boundary between the adsorbed and mixed phases we have rigorous bounds.
We have used Monte Carlo methods to map out the details of the phase diagram and locate this 
fourth phase boundary.
The Monte Carlo results clearly indicate that the boundary between the adsorbed and mixed phases occurs at $y=1$, i.e.~at zero force, unlike the three dimensional case. 
For the ballistic-mixed transition our numerical results match the rigorous results in the asymptotic regime and match results from exact enumeration methods in the non-asymptotic regime.

We have also investigated the nature of the various phase transitions. 
The free-ballistic and free-adsorbed transitions are continuous, consistent with the case of 3-stars in three dimensions and self-avoiding walks, while the ballistic-mixed transition is first order, also like 3-stars in three dimensions.
In two dimensions, the screening of one arm by the adsorption of the other two manifests as both a different location of the phase transition and as a continuous transition rather than the first-order transition in three dimensions.
This reflects the fact that, in two dimensions, the adsorbed phase corresponds to two arms being adsorbed while the third arm is free.
In the mixed phase this third arm becomes ballistic without desorption of the other arms as soon as the pulling force is non-zero.

\section*{Acknowledgement}
C.~B.~thanks York University, Toronto for hosting while part of this work was carried out, 
as well as funding from the University of Melbourne's ECR Global Mobility Award.  
Financial support from NSERC of Canada (Discovery Grant RGPIN-2014-04731) 
and from the Australian Research Council via its Discovery Projects scheme 
(DP160103562) is gratefully acknowledged by the authors.

\end{document}

%% file: figure01.tex
\beginpicture
\setcoordinatesystem units <1.5pt,1.5pt>
\setplotarea x from -30 to 100, y from  0 to 105
\setplotarea x from 0 to 100, y from  0 to 100

\setcoordinatesystem units <1.5pt,1.5pt> point at -40 0 
\color{Tan}
\setplotsymbol ({\Large$\bullet$})
\plot -30 0 130 0  /

\color{red}
\setplotsymbol ({\LARGE.})
\arrow <10pt> [.2,.67] from 40 80 to 40 100
\put {\LARGE$F$} at 47 90

\setplotsymbol ({\large$\bullet$})
\color{RoyalBlue}
\setquadratic
\plot 50 50 40 20 30 22 20 24 10 0  /
\plot 50 50 30 45 20 60 20 70 25 75 35 76 40 80 /
\plot 50 50 55 45 60 25  65 8 70 3 75 6 80 8 85 8 90 5 95 3 100 7  /

\color{Sepia}
\multiput {\scalebox{1.5}{{\huge$\bullet$}}} at 50 50 /
\color{black}
\multiput {\scalebox{1.25}{{\huge$\bullet$}}} at 10 0 /
\color{blue}
\multiput {{\huge$\bullet$}} at 40 80 100 7 /

\color{black}
\normalcolor

\endpicture